# Two-dimensional magnetic tunnel *p-n* junctions for low-power electronics

Wenkai Zhu[1,2], Ziao Wang[1,2], Tiangui Hu[2,3], Zakhar R. Kudrynskyi[4], Tong Zhou[5,6], Zakhar D. Kovalyuk[7], Ce Hu[1,2], Hailong Lin[1,2], Xiaodong Li[2,3], Yongcheng Deng[1,2], Quanshan Lv[1,2], Lixia Zhao[1,3], Amalia Patanè[8], Igor Žutić[5], Houzhi Zheng[1,2], Kaiyou Wang[1,2*]

[1]State Key Laboratory of Semiconductor Physics and Chip Technologies, Institute of Semiconductors, Chinese Academy of Sciences, Beijing 100083, China.

[2]College of Materials Science and Opto-Electronic Technology, University of Chinese Academy of Sciences, Beijing 100049, China.

[3]Tianjin Key Laboratory of Intelligent Control of Electrical Equipment, TianGong University, Tianjin 300387, China.

[4]Advanced Materials Research Group, Faculty of Engineering, University of Nottingham, Nottingham NG7 2RD, UK.

[5]Department of Physics, University at Buffalo, SUNY, Buffalo, NY, USA.

[6]Eastern Institute for Advanced Study, Eastern Institute of Technology, Ningbo, Zhejiang 315200, China.

[7]Frantsevich Institute for Problems of Materials Science, The National Academy of Sciences of Ukraine, Chernivtsi Branch, Chernivtsi 58001, Ukraine.

[8]School of Physics and Astronomy, University of Nottingham, Nottingham NG7 2RD, UK.

*Corresponding author. E-mail: kywang@semi.ac.cn

**For decades, semiconductors and their heterostructures have underpinned both fundamental and applied research across all areas of electronics. Two-dimensional, 2D (atomically thin) semiconductors have now the potential to push further the miniaturization of electronic components, enabling the development of more efficient electronics. Here, we report on a giant anomalous zero-bias spin voltage in magnetic tunnel junctions based on 2D materials. The generation, manipulation and detection of electron spin across a nanometer-thick magnetic tunnel junction do not require any applied bias. It is achieved by exploiting high-quality ferromagnetic/semiconductor interfaces and the asymmetric diffusion of spin-up/spin-down electrons across a semiconductor *p-n* junction. The large spin-voltage signal exceeds 30,000% and is far greater than the highest magnetoresistance signals reported to date. Our findings reveal unexplored opportunities to transform and amplify spin information for low-power electronics.**

The success of charge-based applications in semiconductor heterostructures[1] suggests intriguing opportunities to combine spin and charge degrees of freedom in spintronics[2,3]. To date, key progress in spintronics have focused on spin valves, including magnetic tunnel junctions with metallic ferromagnets. The resistance of these junctions varies with the relative magnetization orientation of the two ferromagnets[2-6]. While metallic spin valves remain valuable for magnetically storing and sensing information, they are of limited use for transferring and processing information, where semiconductors excel. Typically, there are two ways to integrate spin with the versatile charge control in semiconductors: chemical doping and spin injection[2,3,7,8]. Adding magnetic impurities to common semiconductors comes at a high cost: the carrier mobility is drastically reduced and robust emission of light is lost. Alternatively, heterostructures combining common ferromagnets and semiconductors encounter different challenges: spin injection from a ferromagnet into a semiconductor can be strongly suppressed by the interface quality and the resistance mismatch between the materials[2,3]. Although these problems can be alleviated by inserting a tunnel barrier[2,3], the current flow is impeded, requiring a larger applied bias and increased power consumption.

Building on advances in van der Waals (vdW) layered materials and vdW heterostructures[9,10], including a growing number of two-dimensional (2D) magnets[11,12], here we demonstrate a different platform to overcome these obstacles in spintronics. A hallmark of vdW layered materials is their strong covalent atomic bonding within the 2D planes and weak vdW bonding between the layers. This allows for the fabrication of stable thin films down to the monolayer level[10], avoiding interface and crystal defects that are detrimental for spin transport in covalently-bonded magnetic multilayers[13]. As vdW bonding supports atomically-sharp interfaces between dissimilar materials, the same platform can also be suitable for atomically-thin electronic and photonic devices, relying on high-quality metal/semiconductor contacts[14-16]. Here, we report on a giant anomalous zero-bias spin voltage in magnetic tunnel junctions based on 2D magnet and semiconductor *p-n* junctions. Without applying an external bias, the generation, manipulation and detection of electron spin across a nanometer-thick magnetic tunnel junction (MTJ) are achieved by exploiting high quality ferromagnetic/semiconductor interfaces and semiconductor *p-n* junctions.

## Magnetic tunnel *p-n* junctions with anomalous zero-bias spin-voltage

For our vdW heterostructures, as depicted in **Fig. 1a**, we select the layered ferromagnetic metal Fe$_3$GeTe$_2$ (FGT) duo to its desirable large coercivity, strong perpendicular magnetic anisotropy, and relatively high Curie temperature, $T_C$ of up to 220 K[17,18]. Unlike common spin valves, where two ferromagnets are separated by a single nonmagnetic region[2,3,19], here we employ a vdW *p*-GaSe/*n*-InSe heterostructure composed of materials with different doping and composition. The vdW heterostructure is assembled by stacking 2D materials (FGT, γ-InSe, and ε-GaSe) exfoliated from bulk crystals (see Methods for details). The optical image of a typical device is shown in **Fig. 1b**, where the contours of different nanoflakes are outlined by dotted lines of different colors. **Figure 1c** shows the cross-sectional high-resolution transmission electron microscopy (HRTEM) image for a typical device structure. The HRTEM and energy-dispersive X-ray spectroscopy (see **Supplementary Note 1** and **Supplementary Fig. 1**) confirm atomically-sharp interfaces without contamination or amorphous oxide.

Similar to conventional spin-valve devices, passing a charge current *I* through the FGT/*p*-GaSe/*n*-InSe/FGT vdW heterostructure under a magnetic field *B* perpendicular to the layers leads to a magnetoresistance, MR = $|(R_{AP} - R_P)|/R_P$, where $R_{AP}$ ($R_P$) is the resistance for the antiparallel (parallel) configuration of the magnetization in the two FGTs. However, the inclusion of the vdW *p-n* junction leads to significant differences in the measured spin-voltage effect

$$\text{SVE} = |(V_{AP} - V_P)|/V_P, \tag{1}$$

where $V_{AP}$ ($V_P$) is the voltage for antiparallel (parallel) configuration. At a finite current *I*, the SVE coincides with the MR for devices with symmetric linear *I-V* since $V_{AP,P}/I = R_{AP,P}$. Remarkably, we observe a giant and reproducible zero-bias SVE across the *p-n* junction in the absence of any bias current. As shown in **Fig. 1d**, for a device with a 5 nm *p*-GaSe/5 nm *n*-InSe junction (device A), the SVE reaches a value of ~1930% at 10 K. The *V-B* curve exhibits typical SVE with two states. Upon sweeping *B* from negative to positive values, a sharp jump to the high-voltage state occurs at *B* ~0.15 T corresponding to the *AP* configuration ↓↑ or ↑↓, where the arrows denote the out-of-plane FGT magnetization. At *B* ~0.21 T, a sharp jump to the low-voltage state corresponds to the *P* configuration ↓↓ or ↑↑. The voltage jumps are also observed for *B* < 0.

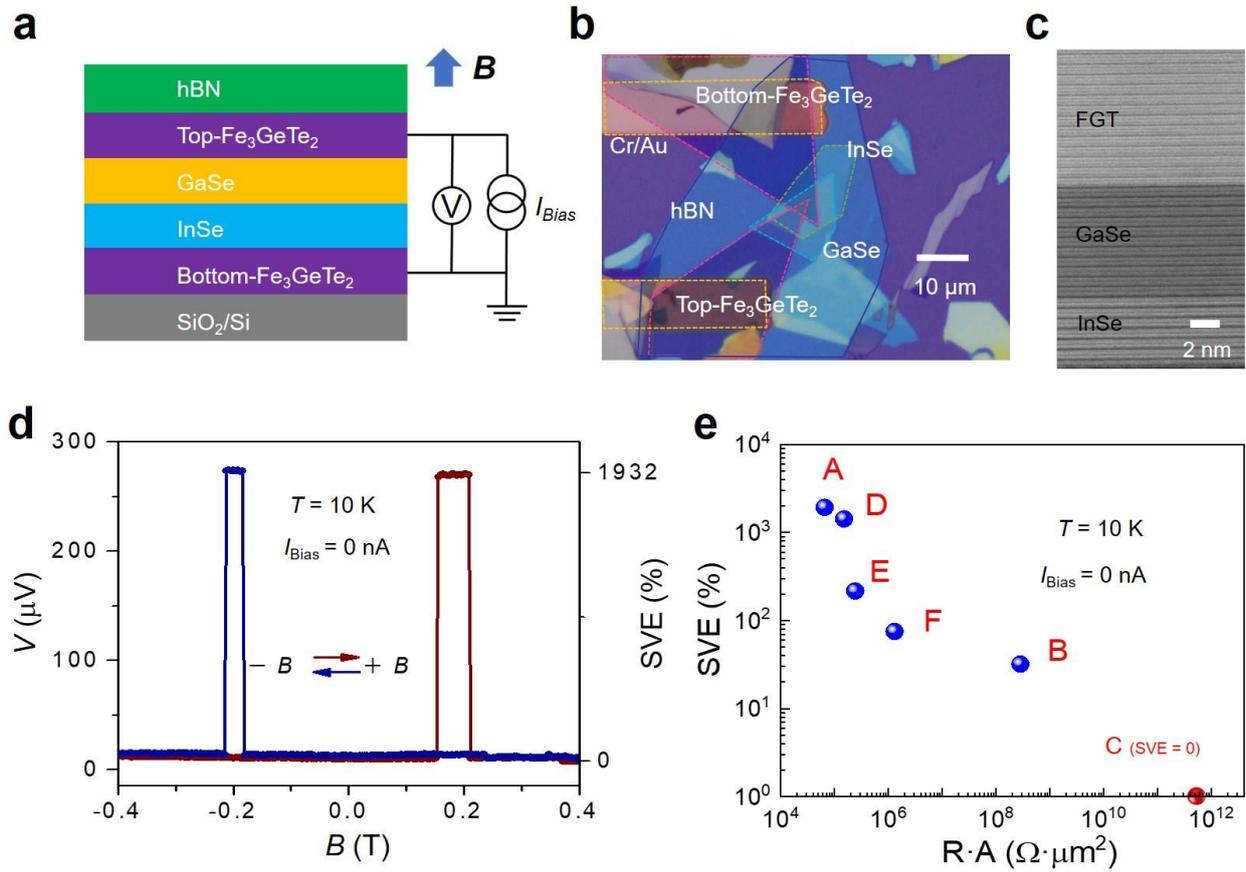

**Fig. 1 | Zero-bias anomalous spin-voltage effect (SVE) in FGT/*p*-GaSe/*n*-InSe/FGT magnetic tunnel junctions. a**, Schematic of the structure with source and drain contacts to the FGT layers and applied perpendicular magnetic field, ***B***. **b**, An optical image of a typical device. **c**, Cross-sectional HRTEM image. **d**, *V-B* curve without external applied voltage for device A shows zero-bias SVE ~1930% at 10 K. **e**, The zero-bias anomalous SVE as a function of the resistance-area product at 10 K for different devices labelled A, B, C, D, E, F.

While a spin signal between two ferromagnets typically increase as their separation decreases[2,3], we observe the opposite behavior when we replace the *p-n* junction with individual thinner *p*- or *n*-type layers. Both MR and SVE are significantly enhanced in the 10 nm thick *p-n* junction compared to devices with a nonmagnetic region of either 5 nm-thick *p*-GaSe or *n*-InSe layers. For devices based on thin individual *p*- or *n*-doped vdW regions, there is no zero-bias SVE (SVE = 0 at $I$ = 0). Even under $I$ = 10 nA, the SVE for *p*-GaSe (~76%)[20] and *n*-InSe (~41%)[21] is significantly smaller than that for the *p*-GaSe/*n*-InSe junction (~258%) (**Supplementary Note 2** and **Supplementary Fig. 2**). This dependence of the spin signal on the thickness of the spacer suggests that our observations do not arise from proximity effects, as proposed in other vdW spin valves[19]. To further explore the dependence of the SVE on the thickness of the *p*-GaSe/*n*-InSe spacer, we investigated a series of *p-n* junctions with a 5 nm-thick *n*-InSe layer and a *p*-GaSe layer of thickness equal to 5, 10 and 15 nm (**Supplementary Note 3** and **Supplementary Figs. 3-4**). The zero-bias SVE decreases with increasing the GaSe thickness from 5 nm (device A, **Fig. 1d**) to 10 nm (device B, **Supplementary Note 3** and **Supplementary Fig. 4a**) and vanishes for 15 nm (device C, **Supplementary Note 3** and **Supplementary Fig. 4b**). As shown in **Fig. 1e**, the zero-bias SVE decreases with increasing the resistance-area product. Thus, the data from different device structures suggest that the observation of the anomalous zero-bias SVE requires a thin *p-n* junction sandwiched between two FGT electrodes.

**Spin-voltage: Temperature and current-dependence**

We now focus on the temperature- and current-dependent behavior of these devices. As shown in **Fig. 2a** for device A, clear *V-B* hysteresis loops without any applied bias current are observed up to the Curie temperature of FGT, $T_c = 220$ K. The measured width of the voltage jumps for the antiparallel magnetic configurations of the FGT electrodes (↓↑ or ↑↓) change non-monotonically with increasing *T*. This is ascribed to the different *T*-dependence of the coercive field in the two FGT layers[22]. **Figure 2b** shows the value of $V_{AP}$ and $V_P$ in the zero-bias state as a function of *T* extracted from the *V-B* curves: unlike the monotonic decrease of $V_{AP}$ with *T*, most likely related to the *T*-dependence of the spin-injection efficiency, the value of $V_P$ varies non-linearly with *T* and changes sign twice. This gives rise to a large SVE and singularities in the SVE versus *T* plot. The corresponding *T*-dependence of the magnitude of the zero-bias spin voltage is shown in **Fig. 2c**. It can be seen that the zero-bias spin voltage reaches a maximum value of 32,230% at $T = 35$K. The corresponding *V-B* loop at $T = 35$ K is presented in **Supplementary Note 4** and **Supplementary Fig. 5**. The non-monotonic temperature dependence of the SVE suggests that, in addition to the properties of the FGT, other mechanisms need to be considered, such as the magnitude and direction of the spin polarization at the FGT/InSe and FGT/GaSe interfaces, and the spin transport along the *p-n* junction.

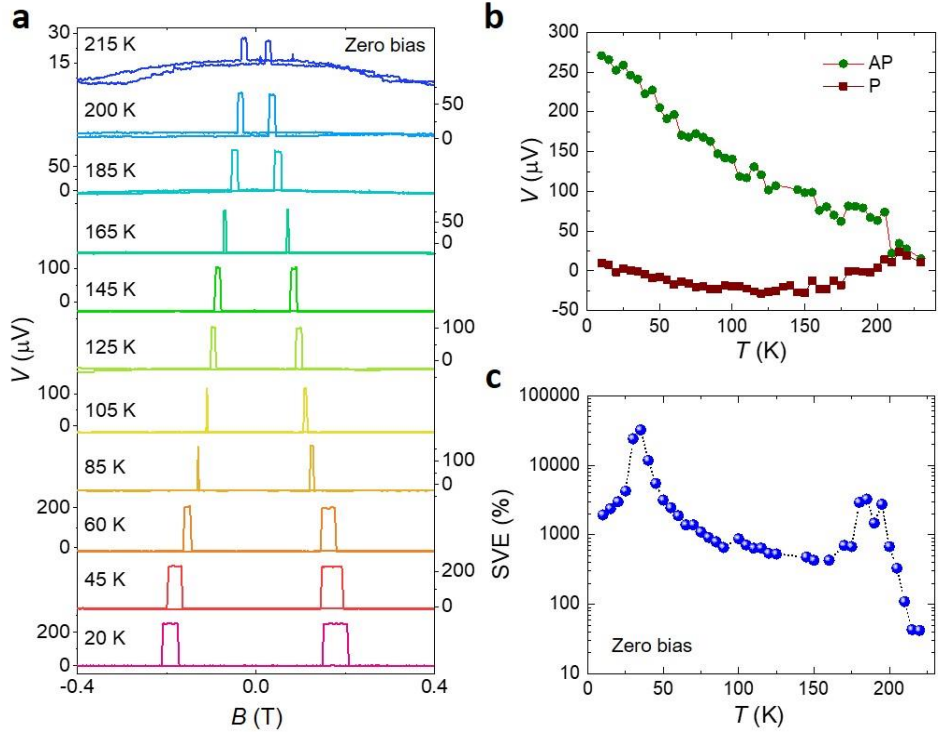

**Fig. 2 | Temperature-dependent zero-bias SVE. a**, Zero-bias SVE for different temperatures $T$ for device A. **b**, The value of $V_{AP}$ and $V_P$ at zero-bias versus $T$, as extracted from the $V$-$B$ curves in part (a). **c**, The zero-bias SVE versus $T$, as derived from the $V$-$B$ curves in part (a).

At a given temperature, the magnitude of the SVE can be optimized by the bias current (**Fig. 3a-d**). For example, at $T = 10$ K, a large SVE of 15,720% is obtained for $I = -1$ nA (**Fig. 3b**). The SVE amplitude extracted from the $V$-$B$ plots (**Fig. 3d**) agrees well with the values obtained from the $I$-$V$ curves (**Fig. 3c**) measured in the parallel and antiparallel configurations. **Figure 3e** shows the color scale plot of the SVE amplitude versus the bias current and temperature. The maximum SVE at low $T$ occurs under a negative current bias, switches to positive current bias at around 40 K, and then switches back under a negative current bias at 180 K, reflecting the $T$-dependence of $V_P$. The conductance $dI/dV$ of the device in the parallel state shows a similar $T$-dependence to that of the bias current for maximum SVE below 170 K (**Fig. 3e**). We attribute this non-monotonic $T$-dependence of the maximum SVE and $dI/dV$ to the competition of two effects: with decreasing $T$ the carrier densities in the $p$-GaSe and $n$-InSe layers decrease, resulting in a smaller built-in electric field; simultaneously the charge/spin scattering rate decreases, leading to longer spin lifetimes. The spin voltage effect in the FGT/$p$-GaSe/$n$-InSe/FGT vertical vdW devices with thin spacer layers is robust and reproducible. The temperature-dependent and bias-dependent SVE for the additional heterostructure device with 5 nm $p$-GaSe/5 nm $n$-InSe junction (device D) are shown in **Supplementary Note 5** and **Supplementary Figs. 6-9**.

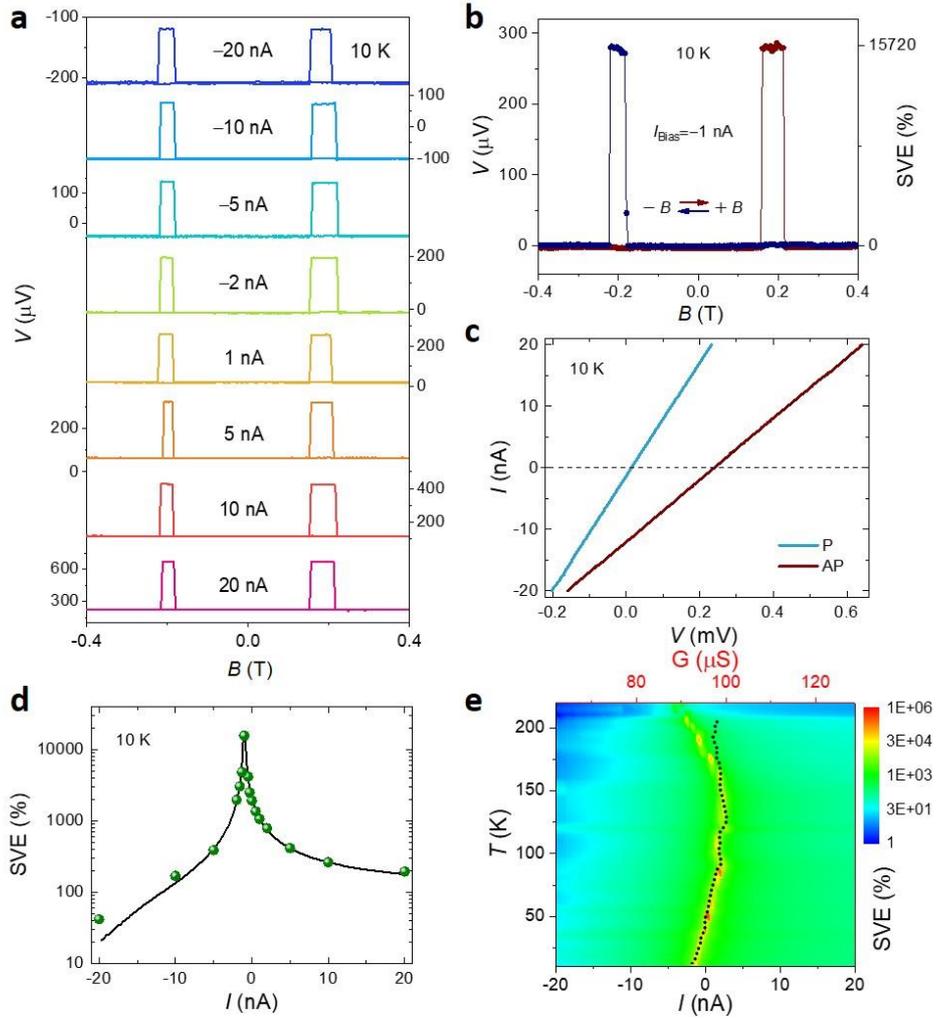

**Fig. 3 | Dependence of SVE on bias current and temperature for device A. a**, $V$-$B$ curves for different bias currents at 10 K. **b**, $V$-$B$ curve under a bias current $I = -1$ nA corresponding to SVE = 15,720%. **c**, $I$-$V$ curves in the parallel and antiparallel magnetization alignments. **d**, SVE as a function of $I$ extracted from the $V$-$B$ (points) and $I$-$V$ (line) curves. **e**, Dependence of SVE on $T$ and $I$. The dotted line shows the conductance versus $T$ in the parallel magnetization alignment.

The zero-bias SVE is not caused by a temperature gradient due to the spin Seebeck effect[23]. To exclude this phenomenon, we designed and fabricated a device structure comprising a magnetic tunnel junction and a heating element. As shown in **Fig. 4a**, the FGT/$p$-GaSe/$n$-InSe/FGT junction (device G) is capped with a few layers of the semi-metal $T_d$-WTe$_2$. This serves as the top heating resistor and is separated from the junction by a few layers of hBN. In this structure, we can create a thermal gradient along the vertical direction by applying an electrical current to the $T_d$-WTe$_2$ layer. The temperature of the $T_d$-WTe$_2$ layer was estimated from its resistance. Firstly, we tested the resistance curve of $T_d$-WTe$_2$ as a function of the heating current. As the heating current increases, the resistance of $T_d$-$WTe_2$ exhibits a quadratic increase, consistent with the Joule heating effect (see **Supplementary Note 6** and **Supplementary Fig. 10**). We examined the relationship between the resistance and temperature of $T_d$-$WTe_2$ under conditions (1 µA) of negligible Joule heating, as illustrated in **Fig. 4b**. Then, the temperature of the $T_d$-$WTe_2$ layer was increased under different heating currents. As the heating current increases from 0 to 0.5, 0.75, and 1 mA, the corresponding temperature of the $T_d$-$WTe_2$ layer increases from 10 to 15, 20, and 25 K, respectively. The hysteresis loops of zero-bias anomalous SVE under different heating currents are shown in **Fig. 4c**. We find that the zero-bias SVE is not sensitive to the thermal gradient. Thus, we conclude that the SVE cannot originate from the generation of a spin voltage due to the thermal injection of spin currents from the FGT into the junction.

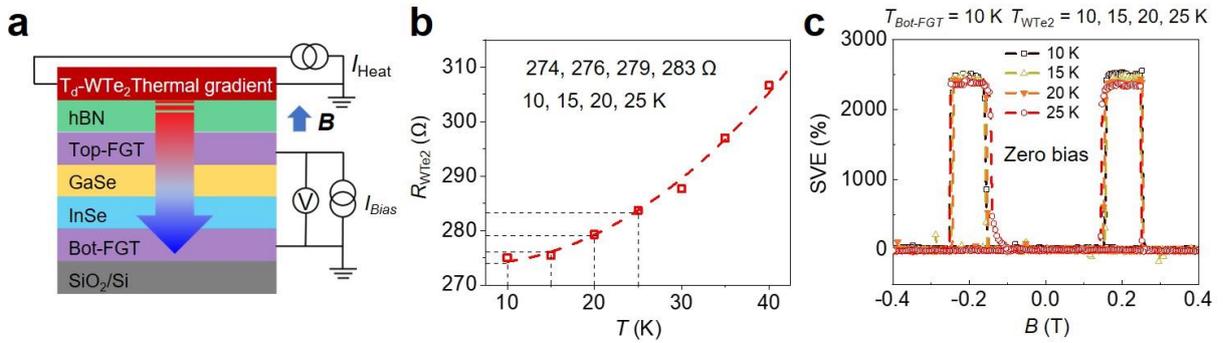

**Fig. 4 | Dependence of the SVE on thermal gradients. a**, The optical image of device G, which consists of a bottom-FGT/5-nm-InSe/5-nm-GaSe/top-FGT/20-nm-$T_d$-$WTe_2$ heterojunction. Joule heating is generated by applying a constant current through the top $T_d$-$WTe_2$ layer of the device, creating a vertical thermal gradient. **b**, The dependence of $T_d$-$WTe_2$ resistance on temperature ranging from 10 K to 40 K at 1 µA. **c**, The SVE curves under different vertical thermal gradients.

We note that the zero-bias SVE effect was observed in several devices with a similar structure, e.g. with a *p-n* junction, supporting the universality of the measured phenomenon. To further test the SVE at zero-bias, we conducted additional experiments: a resistor was connected in series with a magnetic tunnel junction, which acts as a current source in the antiparallel configuration of the FGT (see **Supplementary Note 7** and **Supplementary Fig. 11**).

**Spin-polarized transport in magnetic tunnel junctions**

To explain the zero-bias SVE, we propose a phenomenological model of our magnetic tunnel junction (**Fig. 5a**). For a structure where a single spacer layer (GaSe or InSe) is embedded between two FGT layers (**Fig. 5ai-ii**), the bidirectional diffusion of spins between the FGT cannot generate any spin-voltage at zero-bias, regardless of whether the two FGT have the same or opposite spin polarization orientations. However, when a *p-n* junction is embedded between the FGT layers, the spin diffusion becomes asymmetric due to the built-in electric field of the *p-n* junction (**Fig. 5aiii-iv**). This asymmetry leads to a net spin diffusion in one direction. Using first-principles density functional theory (DFT), we calculate the band structure, electron affinity energy and work-function of ε-GaSe, γ-InSe and FGT. From these calculations, we derive the energy band diagram and electric field of the *p-n* junction (see details in the **Supplementary Notes 8, 9** and **Supplementary Figs. 12-15**).

The spin-polarized transport across the *p-n* junction is simulated using drift-diffusion equations for electrons and holes[24]. Due to the lighter mass and higher mobility of electrons, the spin signal is predominantly determined by electrons (see details in the **Supplementary Note 10** and **Supplementary Figs. 16-19**)[25]. **Figure 5b** shows the spin polarization $P = (n_\uparrow - n_\downarrow)/(n_\uparrow + n_\downarrow)$, where $n_\uparrow$ and $n_\downarrow$ are the spin-resolved electron densities. At the FGT/GaSe interface, we assume $P = 0.66$ at zero-bias voltage, as informed by studies of FGT/hBN/FGT devices[26]. In conventional spin valves, the spatial dependence of the spin polarization $P$ and the spin density $s = n_\uparrow - n_\downarrow$ are equivalent[2,3]. However, while $P$ decays monotonically away from the point of injection of electron spins at the FGT/GaSe interface, the dependence of $s$ is non-monotonic (see inset of **Fig. 5b**). This non-monotonicity induces an asymmetric distribution of spin-polarized electrons at the FGT/GaSe and FGT/InSe interfaces. Furthermore, the measured spin-voltage difference at zero bias between the parallel and antiparallel states is estimated by considering the energy cost of spin flipping at the FGT/InSe interface (see details in the **Supplementary Note 11**). Taking into account the spin polarization of the injected electrons at the interface between GaSe and FGT, the energy difference $\Delta\mu$ at the edge of the InSe is 1.17 meV for *p*-GaSe (5 nm)/*n*-InSe (5 nm) and 0.80 meV for *p*-GaSe (9 nm)/*n*-InSe (5 nm), respectively. For simplicity, the spin splitting of the top and bottom FGT is assumed to be the same, which is consistent with the DFT calculations. Then the energy level diagram for parallel and antiparallel states can be drawn as shown in **Figs. 5c, d**, respectively. The measurable voltage difference between antiparallel and parallel states generated by the spin flipping at the InSe/FGT interface is estimated to be 690.3 μV, which is the same order of magnitude as the experimental observation (see details in the **Supplementary Note 11**). Since the SVE depends on the built-in electric field of the junction as well as on the spin polarization of the FGT/GaSe(InSe) interface, the non-monotonic SVE effect with both temperature and electric field can be explained.

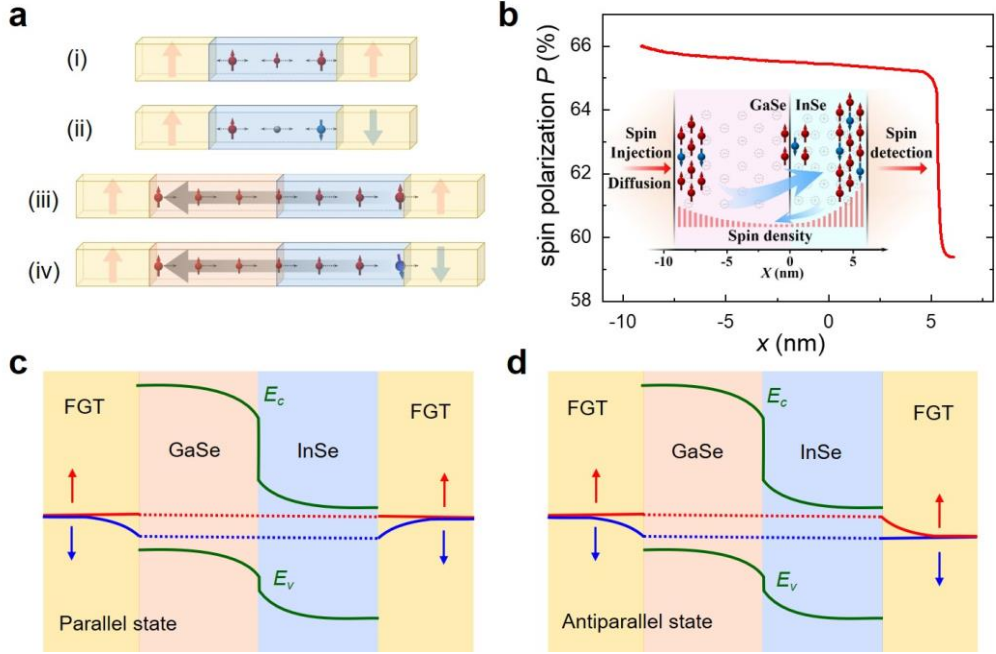

**Fig. 5 | Spin-polarized transport across a magnetic tunnel junction a**, Schematics depicting a single spacer layer between two FGT electrodes in parallel (i) and antiparallel (ii) state, and a *p-n* junction spacer layer between two FGT electrodes in parallel (iii) and antiparallel (iv) state. **b**, Calculated profile of the spin polarization $P$ across the GaSe (5 nm)/InSe (5 nm) junction. Inset: Sketch of the spin density in the junction giving rise to the SVE. **c-d,** Schematic energy band diagram for the FGT/GaSe/InSe/FGT junction in **c**) the parallel state and **d**) antiparallel state. The dotted lines indicate the electron spin up (red) and spin down (blue) quasi-Fermi levels.

Finally, in **Table 1**, we compare the zero-bias SVE signal of our all vdW FGT/*p*-GaSe/*n*-InSe/FGT MTJs with the non-zero or zero bias TMR signal of several other types of MTJs, such as traditional covalent bonding materials-based MTJs, vdW barrier-based MTJs and all vdW MTJs. At low temperature ($T$ = 5 K), MTJs based on the half-metal $CrO_2$/$SnO_2$/Co heterostructure exhibited a zero-bias SVE of 14% that remained without decay for at least 30 minutes[27]. This was ascribed to the diffusion of the spins and their flipping under a strong exchange field[27], leading to an electrochemical potential difference between spin-up and spin-down electrons[28,29]. This difference was detected in early spin-injection experiments[30]. Additionally, a zero-bias SVE of -70% was observed in Co/C/MgO/C/Co MTJ devices, in which the energy source of this spin-driven electrical power is attributed to the harvesting of thermal fluctuations energy[31,32]. As shown in **Table 1**, the TMR magnitude in our all vdW *p-n* junction spin devices is significantly higher, indicating opportunities for higher signal-to-noise ratio and lower energy consumption in spintronic applications.

**Table 1**. Comparison of SVE signal in our all vdW FGT/*p*-GaSe/*n*-InSe/FGT junctions with TMR in other traditional, vdW barrier, and all vdW MTJ devices (* represents the TMR derived from $(dV_{AP}/dI_{AP} - dV_P/dI_P)/dV_P/dI_P$ at zero bias).

|   | Structure | TMR(SVE) | Bias | Temp. | Ref. |
|---|---|---|---|---|---|
| Traditional MTJs | Fe/MgO/Fe | 300% | N/A | 4 K | 5 |
|  | Fe/MgO/Fe/IrMn | 247% | 10 mV | 20 K | 6 |
|  | CoFeB/MgO/CoFeB | 1144% | N/A | 5 K | 33 |
|  | Co/C/MgO/C/Co | -70% | 0 mV | 295 K | 31 |
|  | $CrO_2$/$SnO_2$/Co | 14% | 0 V (*) | 5 K | 27 |
|  | Fe/$MgAl_2O_4$ spinel/Fe | 165% | 0 V (*) | 15 K | 34 |
| vdW barrier MTJs | Co/Graphene/NiFe | 1% | 0 V (*) | 10 K | 35 |
|  | Py/$MoS_2$/Au/Py/Co | 0.73% | 0 V (*) | 20 K | 36 |
|  | Co/hBN/hBN/Co | 12% | 10 mV | 1.4 K | 37 |
|  | Co/hBN/Fe | 50% | 2 mV | 1.4 K | 37 |
| All vdW MTJs | FGT/hBN/FGT | 300% | a.c. 1 mV | 4.2 K | 26,38 |
|  | FGT/$WSe_2$/FGT | 110% | a.c. 1 mV | 4.2 K | 38 |
|  | FGT/GaSe/FGT | 192% | 10 mV | 10 K | 20 |
|  | FGT/InSe/FGT | 41% | 0.1 μA | 10 K | 21 |
|  | FGT/$MoSe_2$/FGT | 42% | 10 mV | 10 K | 39 |
|  | FGT/$MoS_2$/FGT | 4.1% | 1 nA | 10 K | 40 |
|  | $Fe_3GaTe_2$/hBN/$Fe_3GaTe_2$ | 250% | 50 nA | 4 K | 41 |
|  | $Fe_3GaTe_2$/$WSe_2$/$Fe_3GaTe_2$ | 210% | 10 mV | 10 K | 42 |
|  | $Fe_3GaTe_2$/$WS_2$/$Fe_3GaTe_2$ | 213% | 10 nA | 10 K | 43 |
|  | $(Fe_{0.8}Co_{0.2})_3GaTe_2$/$WSe_2$/$Fe_3GaTe_2$ | 180% | 1 mV | 2 K | 44 |
|  | FGT/InSe/GaSe/FGT | 32230% | 0 nA | 35 K | This work |

**Conclusion and outlook**

Our demonstration of efficient spin injection and large magnetoresistance in vdW heterostructures has implications across several fields. The inherent nonlinearities in semiconductors and electrical tunability of 2D ferromagnetism offer opportunities for future developments. Just as the nonlinear response of *p-n* junctions was crucial for early transistors[1], the zero-bias anomalous spin-voltage effect could be employed to implement spin logic and amplification in atomically thin vdW *p-n* junctions. The current use of an applied magnetic field in our experiments could be replaced by using with spin-orbit torques[45] to switch the ferromagnet, enabling seamless integration nonvolatile magnetic memory and spin logic. By using a spin-orbit torque and vdW semiconductors with a direct band gap, the angular momentum from injected spin-polarized carriers could be transferred to control the helicity of the emitted light for low-power and long-distance spin-information transfer[46,47]. While the primary focus of spin valves is their magnetoresistance[2-6], changing the relative magnetization orientation also changes the resulting fringing fields. By considering heterostructures of our FGT-based spin valves with superconductors such fringing fields could potentially support elusive spin-triplet superconductivity, offering a versatile platform for a fault-tolerant topological quantum computing[48,49] by creating Majorana states in 2D geometries[50].

**Methods**

**Fabrication of the FGT/*p*-GaSe/*n*-InSe/FGT vertical heterostructure devices. Figure 1a** in the main text shows a schematic illustration of our spin-valve devices. The few-layer $T_d$-WTe$_2$ (from HQ Graphene), FGT (from HQ Graphene), ε-GaSe (from 2D Semiconductors) and γ-InSe (was grown by the Bridgman method at the Frantsevich Institute for Problems of Materials Science, Ukraine) flakes were mechanically exfoliated using adhesive tape from bulk single crystals of $T_d$-WTe$_2$, FGT, ε-GaSe and γ-InSe, respectively. The few-layer FGT, serving as ferromagnetic electrode, was mechanically exfoliated onto a stamp. Then the stamp was adhered to a glass slide to facilitate handling and identification of thin layers by optical microscopy. Subsequently, the target FGT sheet was transferred onto a SiO$_2$/*p*-Si substrate. Using the same method, the InSe sheet was transferred on top of the FGT sheet. After that, the GaSe sheet was transferred on top of the InSe sheet to fabricate a 2D heterojunction. Finally, another exfoliated few-layer FGT sheet was transferred onto the InSe sheet to form the top electrode. To prevent the FGT from oxidizing once the device was exposed to air, a 15 nm-thick hBN layer was used to cap the entire heterostructure stack. The two FGT sheets were contacted with metallic electrodes, which were fabricated on the substrate prior to the transfer using standard photoetching, thermal evaporation and lift-off. The active overlap area of this vertical device is typically around 5 μm$^2$. The thickness of the FGT, InSe and GaSe flakes was determined by atomic force microscopy (Bruker Multimode 8). The typical thickness of the 2D materials used for the vertical heterostructures in this work is between 5-15 nm. All the fabrication processes were carried out in a glove box with a concentration of less than one part per million of water and oxygen.

**Measurements of SVE effect.** A Keithley 2602 current source and a Keithley 2182A nanovoltmeter were used for the magnetoresistance effect measurements. An Agilent Technology B1500A parameter analyzer was used to study the output characteristics of the spin-valve devices. All measurements were conducted in a Cryogenic Probe Station with magnetic field out-of-plane and at temperatures below and above the Curie temperature ($T_c$ = 220 K) of FGT.

**Computational methods.** The electronic structures of different layers γ-InSe, ε-GaSe, FGT and their heterostructures were investigated using the general potential linearized augmented plane-wave (LAPW) method as implemented in the WIEN2K code. The local density approximation (LDA) was used to describe the exchange and correlation functional, which can yield reasonable electronic structures for InSe, GaSe, and FGT. By fully relaxing both the lattice constants and atomic positions, the optimized lattice constants (a, c) of (3.7 Å 15.7 Å), (4.0 Å, 24.4 Å), and (3.9 Å, 16.3 Å) were obtained for the bulk GaSe, InSe and FGT, respectively. The lattice constants of few-layer films were kept the same as those in the optimized bulk structures.

**Data availability**

The data that support the findings of this study are available from the corresponding author upon reasonable request.

**Code availability**

The codes that support the theoretical part of this study are also available from the corresponding author upon reasonable request.

**Acknowledgements**

K. W. is grateful to Prof. H. Ohno for the useful discussion. This work was supported by the National Key Research and Development Program of China (Grant Nos. 2022YFA1405100), the National Natural Science Foundation of China (Grant Nos. 12241405, 12174384 and 12404146). Z.R.K. acknowledges funding through a Nottingham Research Fellowship from the University of Nottingham. A.P. acknowledges the European Union's Horizon 2020 research and innovation programme Graphene Flagship Core 3. T. Z. and I. Ž. were supported by U.S. DOE, Office of Science BES, Award No. DE-SC0004890.

**Author contributions**

K.W. conceived the work. W.Z. fabricated the devices, W.Z., Z.W., T.H., C.H., H.L., X.L., Y.D., and Q.L. performed the experiments. W.Z., C.H., and K.W. analyzed the data. Z.W., Y.D., I. Ž. and K.W. carried out the modeling. T. Z. and I. Ž. performed DFT calculations; Z.K., Z.K. and A.P. provided the InSe bulk crystals and conducted the preliminary studies of InSe. W.Z., C.H., L.Z., A.P., I.Ž. and K.W wrote the manuscript. All authors discussed the results and commented on the manuscript.

**Competing interests**

The authors declare no competing interests.

**Additional information**

**Supplementary information** is available for this paper.

**Correspondence and requests for materials** should be addressed to K.W.

**Reprints and permissions information** is available at www.nature.com/reprints.

**Online content**

Methods, additional references, Nature Research reporting summaries, source data, extended data, supplementary information, acknowledgements, peer review information; details of author contributions and competing interests; and statements of data and code availability are available.